\documentclass[%
 aip,
 amsmath,amssymb,
 reprint,%
]{revtex4-1}

\usepackage{graphicx}
\usepackage{dcolumn}
\usepackage{bm}

\usepackage[utf8]{inputenc}
\usepackage[T1]{fontenc}
\usepackage{mathptmx}
\usepackage{etoolbox}
\usepackage{comment}
\usepackage{xcolor}

\makeatletter
\def\@email#1#2{%
 \endgroup
 \patchcmd{\titleblock@produce}
  {\frontmatter@RRAPformat}
  {\frontmatter@RRAPformat{\produce@RRAP{*#1\href{mailto:#2}{#2}}}\frontmatter@RRAPformat}
  {}{}
}%
\makeatother
\begin{document}

\preprint{AIP/123-QED}

\title{Flexural Cavity Mechanics in Electrostatically Driven 1D Phononic Crystal}

\author{Vishnu Kumar}
\affiliation{Centre for Nano Science and Engineering, Indian Institute of Science, Bengaluru, 560012, India}
\email{vishnukumar@iisc.ac.in}

\author{Bhargavi B.A.}
\affiliation{Centre for Nano Science and Engineering, Indian Institute of Science, Bengaluru, 560012, India}
\author{Saurabh A. Chandorkar}
\affiliation{Centre for Nano Science and Engineering, Indian Institute of Science, Bengaluru, 560012, India}
\email{saurabhc@iisc.ac.in}


\begin{abstract}
Phononic Crystals provide a versatile platform for controlling phonons in applications such as waveguiding, filtering, and sensing. To minimize dissipation, cavity resonators are often embedded within the bandgap of phononic crystals and integrated with suitable transduction techniques. Here, we demonstrate one-dimensional (1D) phononic transmission using electrostatic transduction, enabling the realization of high-quality mechanical oscillators. Using a double-ended tuning fork resonator embedded in a 1D phononic crystal, we observe degenerate flexural modes (in-phase and out-phase) exhibiting enhanced and comparable quality factors within the same device due to mode degeneracy. The in-phase mode, whose frequency lies inside the phononic bandgap, shows an approximately two-fold increase in quality factor compared to an anchored resonator, while this enhancement diminishes for the out-phase mode (frequency outside the bandgap) at temperatures where thermoelastic dissipation is negligible. This approach offers a promising route toward low-loss, encapsulated phononic devices for sensing and signal processing applications. 
\end{abstract}

\maketitle

Phononic Crystals (PnCs) have recently been getting attention in micro/nano-electromechanical systems (M/NEMS) domains and find numerous applications in acoustic filtering, waveguiding, multiplexing, quantum sensing, bio-sensing and many more \cite{Kumara_phonon_qubit,Mohammadi_HF_Si_Pnc,Zaki_fano,Taleb_ZnO_SAW,Almawgani_review_PnC_sensor,Lucklum_PnC_metamaterials,Pennec_filtering_multiplexing,Casadei_piezo_reso_array}. The periodic arrangements of resonators that obey Bloch conditions form PnCs and facilitate wave propagation and attenuation at specific frequencies, providing solutions to overcome the challenges of mechanical waveguiding. Placing a resonator at the wave attenuated frequencies of the phononic crystal, forming a cavity resonator, effectively confines elastic waves and enhances energy localization, leading to improved quality factors. The achievable quality factor is governed by multiple energy loss mechanisms, including air damping \cite{Bao_Squeeze_film}, thermoelastic dissipation \cite{Lifshitz_TED_NEMS,Chandorkar_Multimode_TED}, anchor damping \cite{Janna_anchor_damping}, Akhiezer damping \cite{Janna_Akhiezer_damping}, as well as piezoelectric and dielectric losses \cite{Vishnu_dielectric,Vish_JMEMS,UCHINO_loss_determination}. The ability of PnC cavities to suppress phonon leakage through anchors has been demonstrated using a variety of architectures, including phononic crystal tethers \cite{Gokhale_PnC_tether,Chen_Four_leaf_AlN,Zhu_Radial_PnC,Awad_Reem_shape_PnC}, crystal strips \cite{Ha_window_PnC_strip}, 2D planer rings \cite{Binci_Planar_PnC}, and honeycomb structures \cite{Schliesser_ultracoherent}. 

While most radio-frequency (RF) applications operate in the megahertz range, realizing phononic crystals in longitudinal modes at these frequencies presents significant challenges in micro and nano-mechanical fabrication \cite{Zivari_nonclassical_PnC,Kim_Topologically_1DPnC,Gomis_1D_optomechanics,Hu_multilayer_nanoparticles,Li_Topological_SSH}. In particular, one-dimensional longitudinal phononic crystals are prone to structural collapse during the release process due to their long periodic chains. To address these fabrication constraints, transverse acoustic modes have been employed for wave propagation and mode localization using diaphragm and beam based architectures.\cite{Chiara_elec_tuning_lattice,Kurosu_temporal_PnC,Kim_bukling_PnC,Kippenberg_SiN_nanobeams,Van_bending_wave_PnC,Marconi_Piezo_array_nrbandgap,Hatanaka_electromechanical_circuits}. Notably, Kurosu et al. demonstrated one-dimensional GaAs/AlGaAs diaphragm-based phononic crystals enabling temporal control of elastic waves and nonlinear phenomena, including phonon soliton generation \cite{Kurosu_temporal_PnC}. 
Since phononic crystals are effective in suppressing anchor damping, integrating them with electrostatic transduction offers a promising route to further enhance device performance, as electrostatic transduction is known to support high quality factors and long-term stability \cite{Saurabh_longterm,Long-term_Stability}. Electrostatically transduced 1D waveguide using suspended graphene on an array of gate electrodes to tune the dispersion has been investigated using the finite element method \cite{Hatanaka_electrostatic_PnC}. Furthermore, V. Zega et al. have investigated the effect of a 2D phononic crystal shield as an acoustic waveguide with two resonators at either end actuated electrostatically and showed improvement in the resonators \cite{Zega_electrostatic_readout}. While these studies demonstrate improved resonator performance, they neither explore the degeneracy mechanisms in double-ended tuning fork (DETF) resonators within a PnC, nor the selective enhancement of quality factors associated with individual flexural modes (in-phase and out-phase).  

In this work, we propose a one-dimensional silicon-based phononic crystal with electrostatic actuation and sensing operating in the transverse flexural mode near the 1 MHz frequency range. Additionally, we have placed a DETF resonator between the phononic crystals to reduce anchor damping. To demonstrate this reduction, thermoelastic damping (TED) was minimized by carrying out the measurement at a temperature at which the thermal coefficient of expansion of silicon tends to zero. The degenerate flexural modes of DETF resonator in a PnC also exhibits their phase modes swapping compared to the anchored DETF resonators.

\begin{figure*}[ht]
    \centering
    \includegraphics[width =0.95\linewidth]{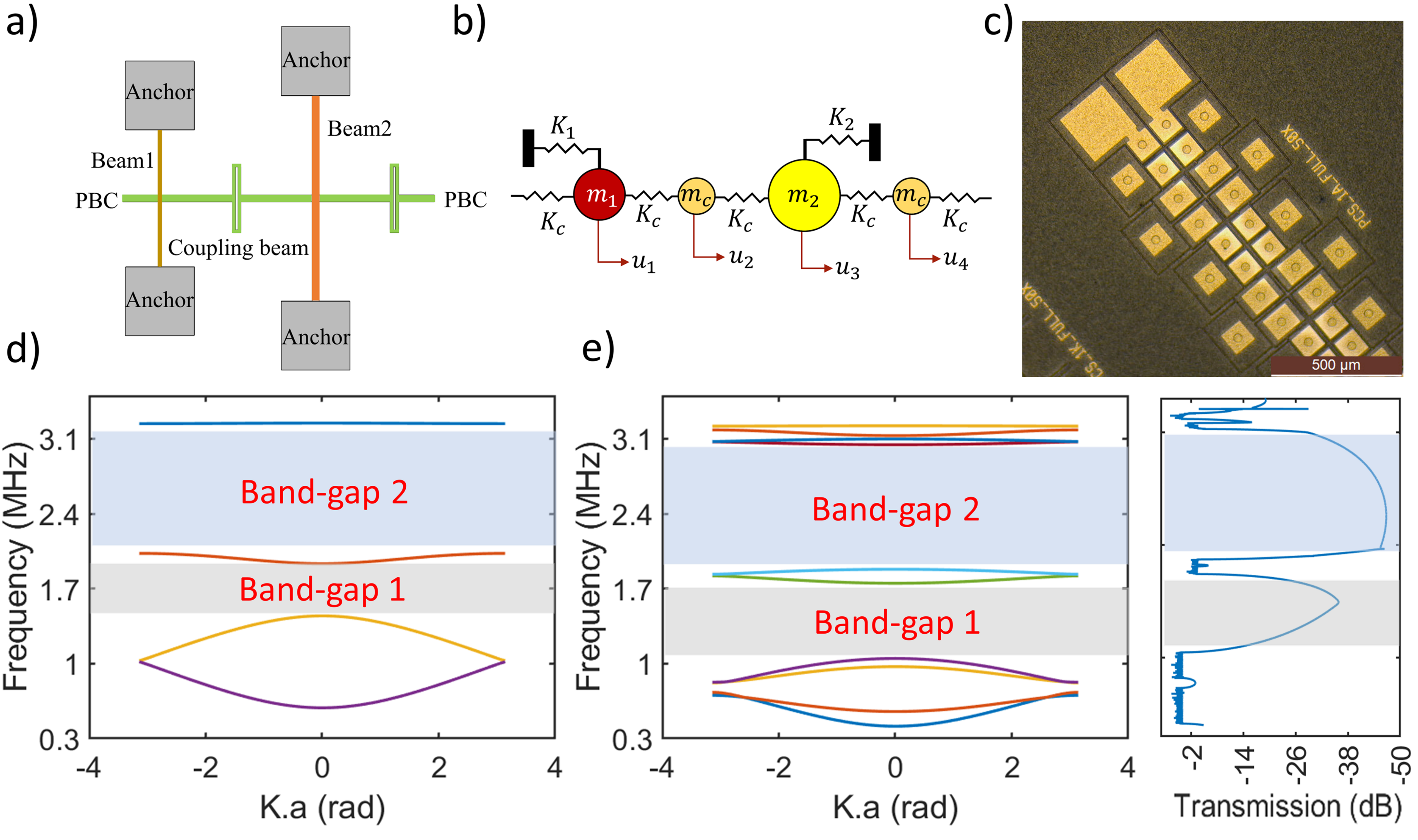}
    \caption{Phononic crystal design: (a) unit cell construction, resonating beam 1 $\&$ 2 coupled with coupling beam, the periodic boundary condition (PBC) is applied in the direction of wave propagation and anchor acts as a fixed boundary condition; (b) spring-mass system configuration; (c) optical image of the PnC device; (d) analytically calculated dispersion curve shows the first two bandgaps; (e) COMSOL simulated dispersion curve and transmission}
    \label{PnC_design}
\end{figure*}

We design and fabricate (refer supplementary section I for the fabrication details) a phononic crystal composed of repeating unit cells consisting of two fixed-fixed beams connected via a coupling beam. Each beam functions as a spring-mass-damper (SMD) system. Fig. \ref{PnC_design}{a} shows the phononic crystal unit cell and corresponding SMD configuration is shown in fig. \ref{PnC_design}{b}. The masses ($m_1$ and $m_2$) represent the two fixed-fixed beams, while the mass ($m_c$) represents the coupling spring connecting the resonating beams. The coupling spring's influence can alter the dynamics of the crystals. To increase the range of bandgaps, the coupling spring's strength is set lower than that of the fixed-fixed beams.
An array of these unit cells, with a total of 25 cells, sufficiently enhances mechanical signal reflections. The optical image of the PnC device is shown in fig. \ref{PnC_design}{c}.

For the phononic unit cell, the equations of motion are given by (neglecting air damping as all the measurements are carried out in a vacuum, and the dispersion is unaffected by the damping of the first order):

\begin{subequations}
\centering
    \begin{gather}
        m_1\Ddot{u_1}^n+K_1u_1^n+K_c(u_1^n-u_2^n)+K_c(u_1^n-u_2^{n-1})=0 \\
        m_c\Ddot{u_2}^n+K_c(u_2^n-u_1^n)+K_c(u_2^n-u_3^n)=0\\
        m_2\Ddot{u_3}^n+K_2u_3^n+K_c(u_3^n-u_2^n)+K_c(u_3^n-u_4^n)=0\\
        m_c\Ddot{u_4}^n+K_c(u_4^n-u_3^n)+K_c(u_4^n-u_1^{n+1})=0
    \end{gather}
\end{subequations}

Applying plane wave solution $u^n=u^0*exp(i(kx-\omega t))$, assuming the length of the unit cell to be $a$, the device dimensions and derived parameters $m_1$, $m_2$, $m_c$, $k_1$, $k_2$, and $k_c$ are shown in the Table \ref{device_parameters}.

\begin{table}[!h]
    \centering
    \begin{tabular}{|c|c|c|c|c|c|} \hline
          $m_1 (Kg)$ & $m_2 (Kg)$ & $m_c (Kg)$ & $k_1 (N/m)$ & $k_2 (N/m)$ & $k_c (N/m)$ \\ \hline
          $2.9e-11$ & $1.276e-10$ & $1.18e-10$ & $1073.6$ & $5767.3$ & $4874.4$ \\ \hline
    \end{tabular}
    \caption{Device parameters used for the dispersion solution}
    \label{device_parameters}
\end{table}

We solved for the eigenvalues at each wave-vector, and the resultant dispersion curve is shown in fig. \ref{PnC_design}{d}. The bandgaps occur in the ranges 1.4 MHz to 1.96 MHz and 2.1 MHz to 3.2 MHz. These two bandgaps result from the interactions between the resonating beams and the coupling structure, which is also part of the resonating system. 

We also performed a FEM simulation by using the COMSOL multiphysics structural mechanics module. The unit cell is designed in a 2D geometry for in-plane vibration. A fixed boundary condition is applied to resonating beams 1 and 2, while the Floquet boundary condition is applied to the two ends of the unit cell. The simulated eigenfrequencies at each wave vector were plotted with some post-processing in Matlab, as shown in fig. \ref{PnC_design}{e}. The bandgap frequency ranges, identified from the dispersion curve, indicate regions where no modes exist. In other words, waves at these frequencies are reflected back, resulting in a dip in transmission.  

The multiple geometrical parameters corresponding to various bandgap formations are detailed in the supplementary section II.

\begin{figure}[ht]
    \centering
    \includegraphics[width =\linewidth]{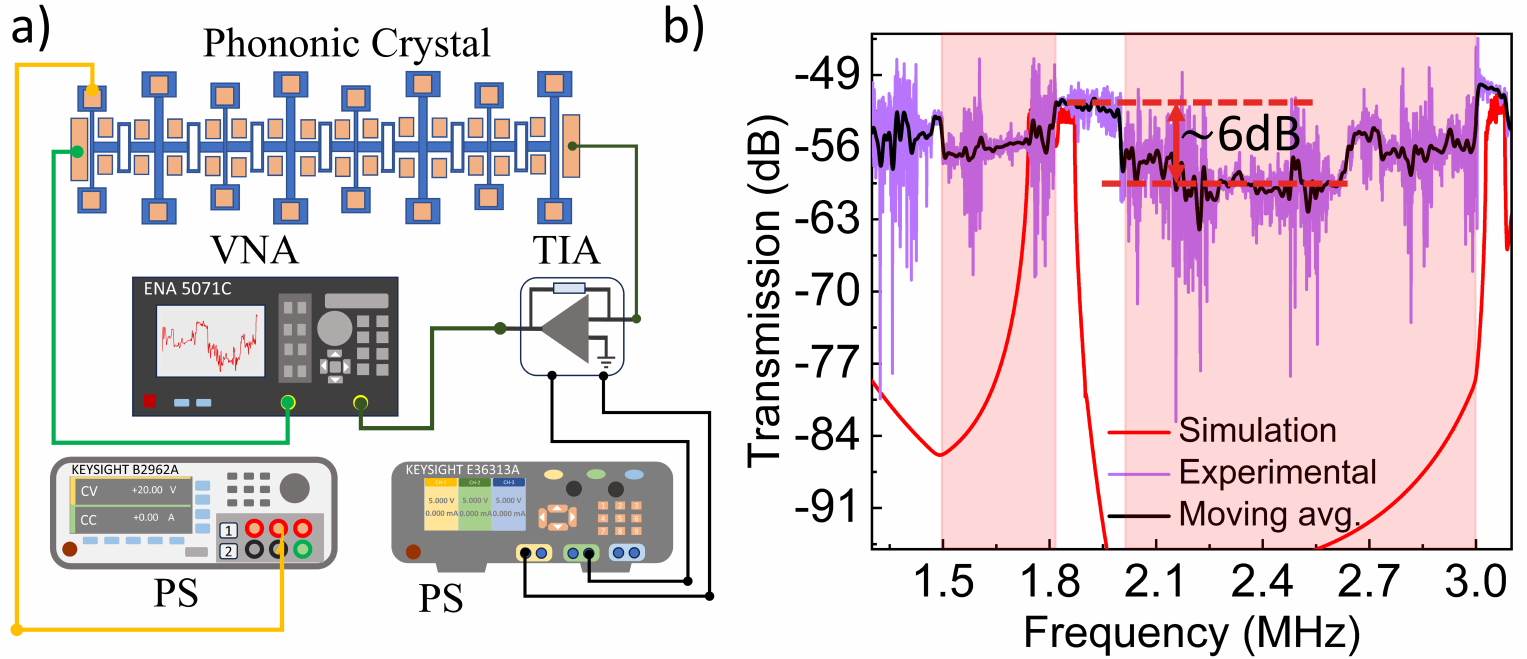}
    \caption{Transmission characterization of PnC: (a) the measurement setup, (b) the measured transmission along with the simulation. VNA: Vector Network Analyzer, TIA: Trans-Impedance Amplifier, and PS: Power Supply}
    \label{PnC_Trans}
\end{figure}

Electrostatic actuation and sensing techniques have been employed to measure the phononic crystals. The in-plane motion is excited by applying an AC signal to the actuation electrode ($V_{in}$), and the resulting motion is measured at the sensing electrode ($i_{out}$), using a network analyzer. At first, to perform the measurements, a Short-Open-Load-Transmission (SOLT) calibration was carried out on the signal transmission cables prior to connecting the network analyzer to the PnC device. Following calibration, a Keysight ENA E5071C was used to apply the actuation signal, while a DC bias was supplied to the PnC using a Keysight B2962A. Upon application of the DC bias voltage, the movable beams within the PnC experience electrostatic forces, inducing mechanical vibrations at the excitation frequency. As multiple unit cells are coupled to form the PnC, the resulting mechanical waves propagate through the structure, undergoing interactions across successive cells before reaching the terminal unit. The output signal is detected by an electrode located at the end cell and subsequently routed to a transimpedance amplifier (TIA). The TIA, powered by a Keysight E36313A with ±5 V bias, converts the current signal into a voltage signal with sufficient gain (refer supplementary section III). The amplified voltage is then fed into the second port of the network analyzer, where the transmission parameter is measured, as shown in Fig. \ref{PnC_Trans}(a).

Transmission measurements were conducted using a phononic crystal (PnC) consisting of 25 unit cells. The PnC was designed with 25 unit cells to achieve close to -40 dB transmission attenuation. During the measurements, the power level was set to -10 dB, with a bias voltage of 40 V. The transmission spectra, along with the simulated result, are depicted in fig. \ref{PnC_Trans}{b}.

The PnC's transmission strength is $\sim$ 6 dB, significantly less than the simulated strength. Several factors have contributed to this discrepancy. One potential cause is the presence of large capacitance resulting from multiple electrodes with floating potentials. Another factor could be fabrication issues; abnormalities during lithography and etching may have caused dimensional shifts in the phononic crystal structure, reducing the overall transmission strength.

\begin{figure}[!h]
    \centering
    \includegraphics[width =\linewidth]{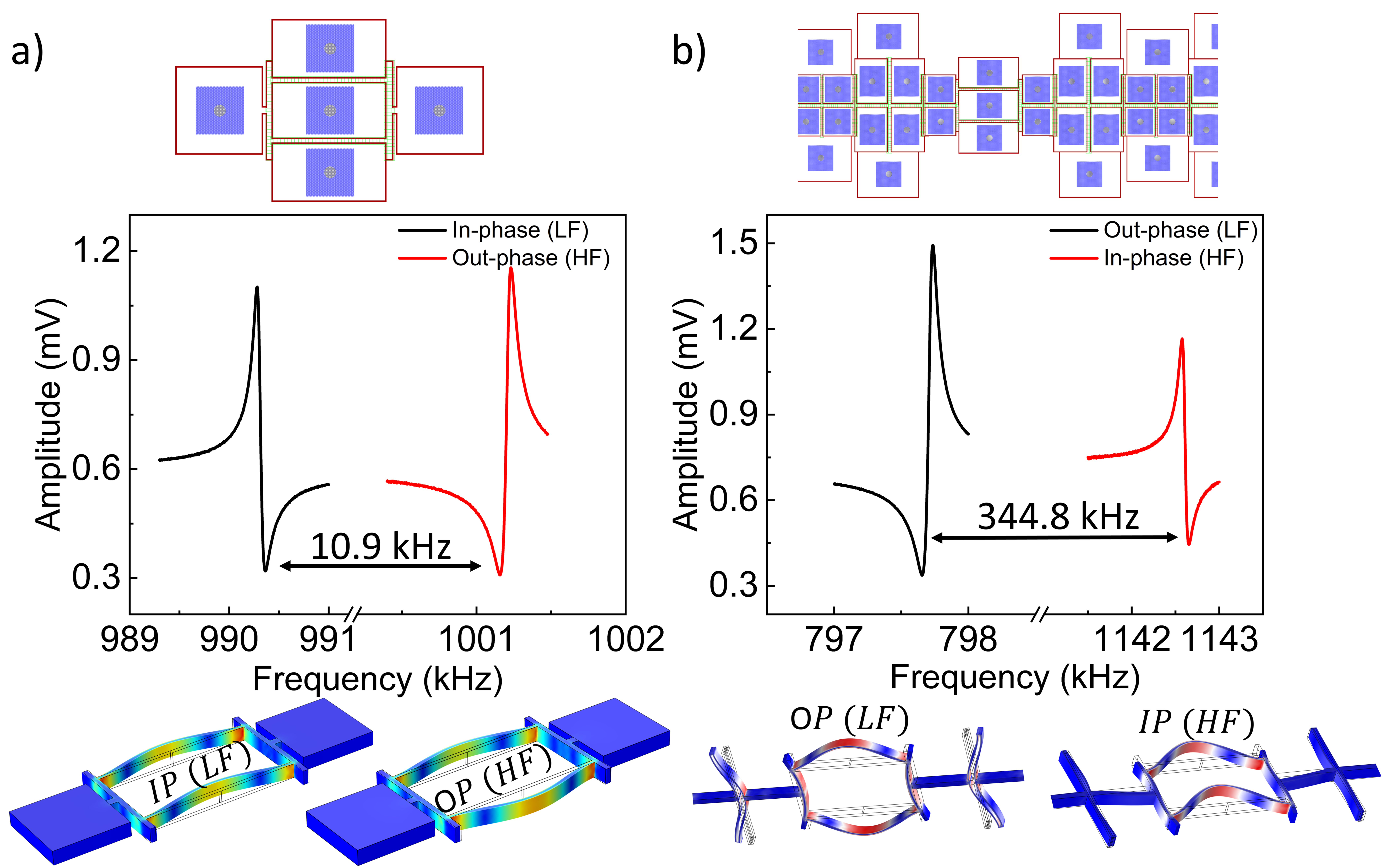}
    \caption{Anchor Resonator: (a) the measurement schematic with the resonator design, (b) The frequency response of the resonator shows two degenerate modes: the low-frequency (in-phase) mode and the high-frequency (out-phase) mode, and COMSOL simulation of the resonator corresponding to the degenerate modes.}
    \label{anchor_cavity_1stmode}
\end{figure}

For our flexural mode cavity mechanics study, we used a double-ended tuning fork (DETF) resonator as a cavity resonator. Initially, we examined the resonator with its anchor. The DETF anchored resonator comprises two resonating beams connected by a coupling beam that is anchored in the middle, as illustrated in fig. \ref{anchor_cavity_1stmode}{a}. These two resonating beams produce degenerate modes through the coupling beam. 

The degenerate first mode of the DETF with the anchor is shown in fig. \ref{anchor_cavity_1stmode}{a}. The frequency of the in-phase (IP) mode is approximately 990.31 kHz, which is lower than the out-of-phase (OP) mode frequency of about 1001.2 kHz. The frequency difference between these two modes indicates the strength of the coupling beams that connect the two resonating beams. The strength can be calculated using the equation of motion of the degenerate modes ($0.5*m_{eff}(\omega_{OP}^2-\omega_{IP}^2=K_c)$). 

All these measurements are conducted in vacuum to minimize the effects of air damping, ensuring that the mechanical properties and material of the device primarily influence the quality factor. Since silicon is used, energy losses will predominantly result from thermoelastic dissipation and anchor damping. The quality factor of the anchored resonator at room temperature is limited by thermoelastic dissipation (TED). The experimental quality factors for the responses shown in fig. \ref{anchor_cavity_1stmode}{a} are $\sim$14922.6 for the low-frequency IP mode and $\sim$17175.6 for the high-frequency OP mode. These experimental Qs are obtained after subtracting the signal at DC bias off to eliminate the background signal coming due to feedthrough. Using the thermoelasticity module in COMSOL to estimate TED, the simulated quality factors are $\sim$17540 for the low-frequency IP mode and $\sim$17487 for the high-frequency OP mode, demonstrating the effect of TED.

A DETF resonator is positioned within the phononic crystal to prevent energy leakage from the anchor. Outer electrodes are used for capacitive actuation and sensing, and one nearby anchored electrode at the phononic site is used for DC biasing. The optical image of the cavity resonator is depicted in the supplementary section I. The output signal from one of the outer electrodes of the resonator is connected to the TIA for resonance response. 

The frequency response of the cavity is shown in fig. \ref{anchor_cavity_1stmode}{b}, where the degenerate modes (IP and OP) of the first fundamental mode are captured. The frequency response in black represents the out-phase vibration mode at $\sim$797.7 kHz, whereas the response in red represents the in-phase vibration mode at $\sim$1142.6 kHz. The degenerate modes are described through COMSOL simulation, shown in fig. \ref{anchor_cavity_1stmode}{b}. The simulated dispersion and transmission results indicate that the measured frequencies are 1) outside the bandgap for the out-phase mode and 2) inside the bandgap for the in-phase mode, as represented in Figure S5 (refer supplementary section IV).

From the simulation, the OP mode is visualized as the combination of the first vibration mode of resonating and coupling beams. The effect of the strain profile can be observed as both beam modes show a thermal gradient along with the coupling spring. This reduces the quality factor of the mode, ( $\sim$13387.9). Meanwhile, the IP mode is a combination of the first mode of resonating beam and the second mode of coupling beam, and the thermal gradient is seen within the resonator. Thus, the quality factor is slightly higher ($\sim$17201.34). 

The frequency separation between the degenerate modes of the cavity resonator ($\sim$ 345 kHz) is significantly greater than that of the anchored resonator ($\sim$ 11 kHz). Also, the in-phase mode vibration is shifted to a higher frequency compared to the lower frequency of the anchored resonator. This phenomenon occurs due to the coupling of the phononic crystal with a resonator. Once the resonator is within the PnC, the stiffness at the point of contact is effectively low (or even zero), suggesting that the resonator's point of contact with the PnC acts as a free boundary. This effect is simulated in COMSOL without applying any constraints to validate the phenomena. The degenerate modes frequency separation is close to the experimental separation, as shown in fig. S7 (refer supplementary section IV).

\begin{figure}[!h]
    \centering
    \includegraphics[width =\linewidth]{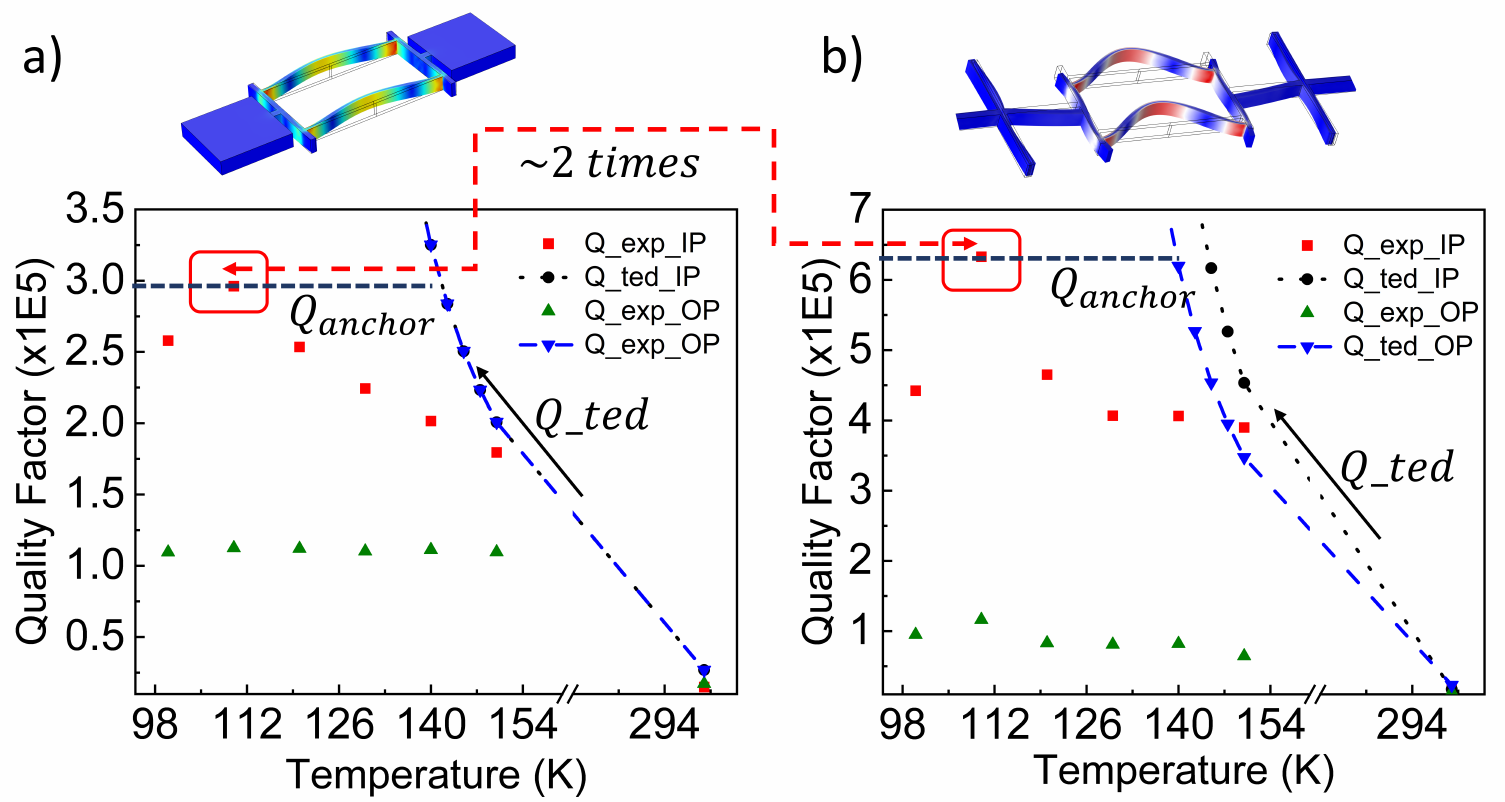}
    \caption{(a) anchored resonator near the silicon CTE temperature, the in-phase mode has a higher quality factor than the out-phase mode, (b) cavity resonator, the in-phase has a higher Q than the out-phase and two times the anchored resonator.}
    \label{compare}
\end{figure}

With the behavior of the two degenerate modes known at room temperature and the quality factor limited by thermoelastic dissipation (TED), we conducted measurements at temperatures where the coefficient of thermal expansion (CTE) of silicon is nearly zero. This was done for both the anchored and cavity resonators. Notably, the quality factor of in-phase mode of cavity resonator has shown approximately a two fold change, indicating a reduction in anchor loss for the DETF within the phononic crystal at a temperature of 110 K, as shown in fig. \ref{compare}. The observed $\sim$ 2 times change is attributed to the 6 dB attenuation in the bandgap region. Simulated TED results show a very high quality factor at this temperature, leaving anchor damping as the primary factor. The out-phase mode for both resonators did not change significantly since this mode lies outside the bandgap. Therefore, it is demonstrated that placing a resonator within phononic crystals and operating at frequencies within the bandgap enhances the quality factor.

We report electrostatically actuated and sensed transmission in a one-dimensional beam-based phononic crystal, revealing multiple bandgaps consistent with numerical and finite element analyses. A double-ended tuning fork resonator embedded within a phononic cavity exhibits pronounced modifications in its degenerate phase modes compared to an anchored resonator. At 110 K, where thermoelastic damping is negligible, the in-phase mode shows an approximately twofold increase in quality factor due to its frequency lying within the bandgap, whereas the out-of-phase mode exhibits only a marginal enhancement ($\sim$1.03 times) as it lies outside the bandgap. These results demonstrate mode-selective dissipation control using phononic crystals for acoustic signal processing applications.

We wish to acknowledge the support of the Centre for Nanoscience and Engineering, IISc, the National Nano Fabrication Centre (NNFC), and the Micro-Nano Characterization Facility (MNCF). Vishnu Kumar gratefully acknowledges the MHRD, Govt. of India, for providing us with the necessary funding and fellowship to pursue research work. Saurabh A. Chandorkar acknowledges Indian Space Research Organization, Government of India (GoI) under Grant DS 2B13012(2)/41/2018-Sec.2, by the Ministry of Electronics and Information Technology, GoI under 25(2)/2020 ESDA DT.28.05.2020 and Ministry of Human Resource and Development, GoI Grant SR/MHRD 18 0017. Saurabh A. Chandorkar also acknowledges DRDO JATP: DFTM/02/3125/M/12/MNSST-03.

V.K. designed, fabricated, performed the measurement, and prepared the manuscript with the help of S.A.C. B.B.A. helped in device fabrication. S.A.C. supervised the whole project.

\bibliography{References}

\end{document}


\title{{\centering Supplementary Material for:\\
Flexural Cavity Mechanics in Electrostatically Driven 1D Phononic Crystal}}

\author{Vishnu Kumar}
\affiliation{Centre for Nano Science and Engineering, Indian Institute of Science, Bengaluru, 560012, India}
\email{saurabhc@iisc.ac.in}
\email{vishnukumar@iisc.ac.in}

\author{Bhargavi B.A.}
\affiliation{Centre for Nano Science and Engineering, Indian Institute of Science, Bengaluru, 560012, India}
\author{Saurabh A. Chandorkar}
\affiliation{Centre for Nano Science and Engineering, Indian Institute of Science, Bengaluru, 560012, India}


\maketitle

\section{Device Fabrication}

\begin{figure}[!h]
    \centering
    \includegraphics[width =\linewidth]{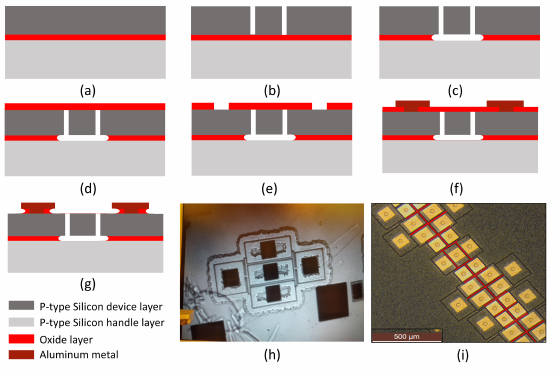}
    \caption{Device fabrication: (a-g) fabrication process steps involved, a) SOI wafer, b) DRIE to etch deep trenches, c) first timed vapor HF to etch box oxide to reduce the dependency of long etch during final device release, d) LTO deposition, e) LTO etch to open the area for metal contact, f) metal deposition and etch to form electrodes, g) timed vapor HF to release the device from the top; h) Infrared imaging from bottom to check the vapor HF etch release, i) the PnC device with the DETF cavity resonator.}
    \label{fab_steps}
\end{figure}

The fabrication process of electrostatically driven phononic crystals starts with the highly boron-doped silicon-on-insulator (SOI) platform having a silicon device layer thickness of 20 $\mu$m and box layer of 2 $\mu$m. The wafer is first RCA cleaned, followed by diluted HF to remove unwanted particles and the native oxide layer. Silicon dioxide of 1 $\mu$m as a hard mask is deposited by using the Plasma Enhanced Chemical Vapor Deposition (PECVD) to help in etching deep trenches. The first layer, using optical lithography, defines the structure; this was achieved by coating the AZ5214E photoresist, followed by UV exposure and development of the sample. The PECVD oxide hard mask is etched using Reactive Ion Etching (RIE), and then the silicon device layer is etched using Deep RIE. Once the trenches are defined, time-dependent vapor HF is performed to etch the box oxide from the bottom of the device silicon. The etch is limited to a few $\mu$m to prevent over-etching in later device release. The sample is then placed in the Low-Pressure Chemical Vapor Deposition (LPCVD) chamber to deposit low-temperature oxide (LTO) at 450$^{\circ}$C of 3 $\mu$m on top of the trenches to prevent metal from entering inside the trenches. The second layer, using a similar optical lithography technique by coating AZ4562 photoresist, is performed, followed by RIE to open the area that connects the device layer to the metal. Next, the Aluminium metal of $\sim$ 4 $\mu$m is deposited by sputtering on the wafer as electrode metal. The third optical lithography layer is performed, followed by a metal etch to pattern the electrodes. After removing the photoresist residual, vapor HF is again performed to release the device from the top; this is due to the etch rate variation of LTO with box oxide (LTO etch rate is faster than the thermal oxide). The device is then inspected in the IR camera from the backside to check the etch release front. The corresponding steps are shown in fig. \ref{fab_steps}.

\section{Phononic Crystals Simulations}

The COMSOL simulation of phononic crystals with varying geometric parameters is presented, focusing on the in-plane motion of the system, as it is electrostatically actuated and sensed in-plane.

\subsection*{Effect of coupling beam}

\begin{figure}[!h]
    \centering
    \includegraphics[width=\linewidth]{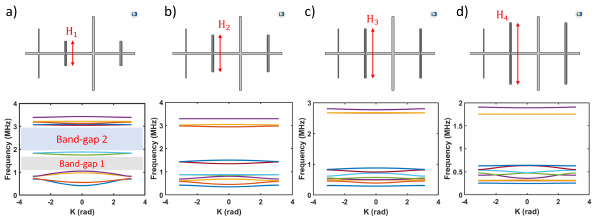}
    \caption{Variation in the serpentine structure length and associated simulated dispersion: a. $H_1 = $100 $\mu $m, b. $H_2 = $150 $\mu $m, c. $H_3 = $200 $\mu $m, and d. $H_4 = $100 $\mu $m}
    \label{H_cb_change}
\end{figure}

First, the coupling beam parameters are modified to study their effect on the PnC dispersion. fig. \ref{H_cb_change} illustrates the bandgap variation resulting from changes in the height of the serpentine section of the coupling beam. In these simulations, the number of vibration modes remains constant across all the simulations for comparison. The results show that the lower bandgap (bandgap 1) gradually disappears, merging with the lower frequency modes, while the upper bandgap (bandgap 2) widens. This adjustment enables the engineering of bandgaps to meet specific requirements, with larger bandgaps offering significant advantages in filtering applications \cite{Pennec_filtering_multiplexing}.

\begin{figure}[!h]
    \centering
    \includegraphics[width=\linewidth]{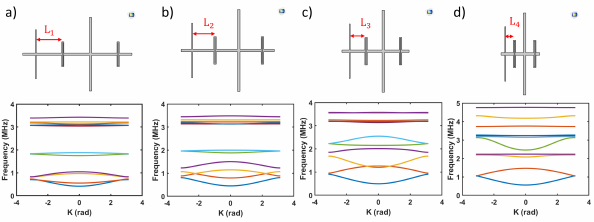}
    \caption{Variation in the length of coupling beam and associated simulated dispersion: a. $L_1 = $100 $\mu $m, b. $L_2 = $75 $\mu $m, c. $L_3 = $50 $\mu $m, and d. $L_4 = $25 $\mu $m}
    \label{L_cb_change}
\end{figure}

Other geometrical parameters, such as the length of the coupling beam, can also significantly influence the phononic dispersion, as variations in length alter the beam’s stiffness, thereby impacting bandgap formation. Fig. \ref{L_cb_change} illustrates the evolution of multiple bandgaps resulting from changes in coupling beam length. Shortening the beam increases stiffness, which, in turn, weakens its interaction with the resonating beams and reduces the bandgap. The 2$^{nd}$ and 3$^{rd}$ modes interaction results in mode overlapping as the length reduces, as seen in fig. \ref{L_cb_change} b $\&$ c. The combination of the different geometrical parameters of the coupling beam enables a wide range of phononic engineering for various applications in filtering, sensing, etc \cite{Pennec_filtering_multiplexing, Almawgani_review_PnC_sensor}. 

\subsection*{Effect of resonating beams}
\begin{figure}[!h]
    \centering
    \includegraphics[width=\linewidth]{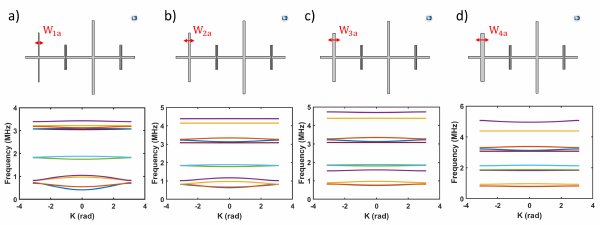}
    \caption{Variation in the width of the resonating beam 1 and associated simulated dispersion: a. $W_{1a} = $4 $\mu $m, b. $W_{2a} = $8 $\mu $m, c. $W_{3a} = $12 $\mu $m, and d. $W_{4a} = $16 $\mu $m}
    \label{W_m1_change}
\end{figure}

Resonating beams 1 and 2 are interconnected by a coupling beam. Altering the geometric parameters of either resonating beam affects mass and stiffness, leading to variations in phononic dispersion. fig. \ref{W_m1_change} demonstrates how adjusting the width of resonating beam 1 ($W_{1a}$, $W_{2a}$, $W_{3a}$, and $W_{4a}$) opens multiple bandgaps, each distinct but narrower. A variation of 4 $\mu $m was applied for comparative analysis. 

\begin{figure}[!h]
    \centering
    \includegraphics[width=\linewidth]{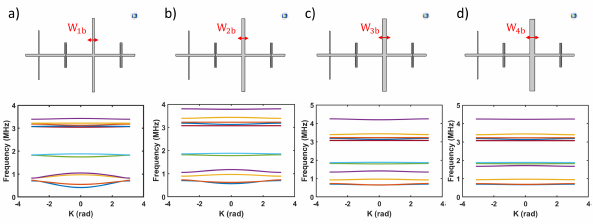}
    \caption{Variation in the width of the resonating beam 2 and associated simulated dispersion: a. $W_{1b} = $10 $\mu $m, b. $W_{2b} = $14 $\mu $m, c. $W_{3b} = $18 $\mu $m, and d. $W_{4b} = $22 $\mu $m}
    \label{W_m2_change}
\end{figure}

Furthermore, varying the width of the resonating beam 2 ($W_{1b}$, $W_{2b}$, $W_{3b}$, and $W_{4b}$), the phononic dispersion as shown in fig. \ref{W_m2_change} displays similar behavior as shown in fig. \ref{W_m1_change}. 

\section{Trans-Impedance design}

\begin{figure}[!h]
    \centering
    \includegraphics[width =\linewidth]{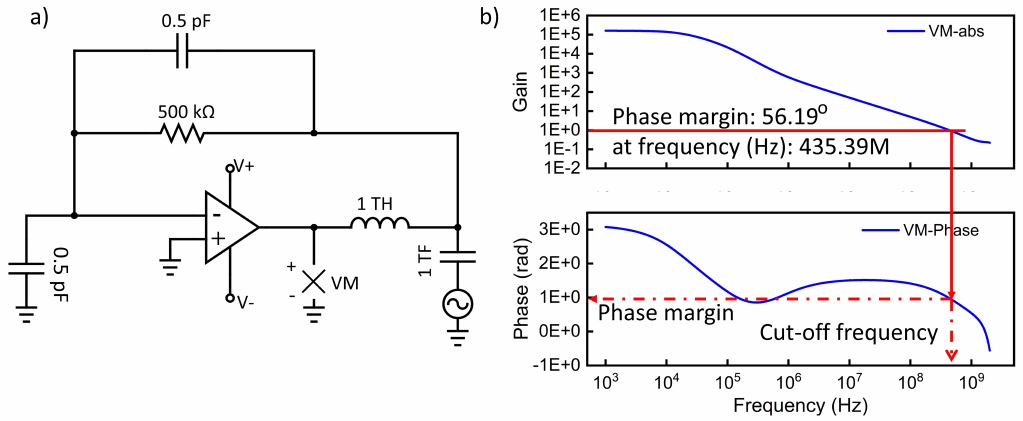}
    \caption{The trans-impedance amplifier design: a) schematic of the TIA for the stability check, b) the response of the TIA circuit (phase margin is \(56^\circ \))}
    \label{TIA}
\end{figure}

The TIA simulation was performed on Texas Instruments' TINA-TI software. In the simulation, OPA 818 opamp is used for the active amplification with feedback consisting of a resistor and capacitor in parallel configuration. To ensure stability, the phase margin of the circuit system should be larger than \( 45^\circ \). With the \( R = 500 \, \text{k}\Omega \) and \(C = 0.5 \, pF\), the phase margin came out to be \(56.19^\circ \), satisfying the stability condition. The simulation is shown in fig. \ref{TIA}.

\section{Effect of PnC coupling on cavity resonator}

\begin{figure}[!h]
    \centering
    \includegraphics[width =\linewidth]{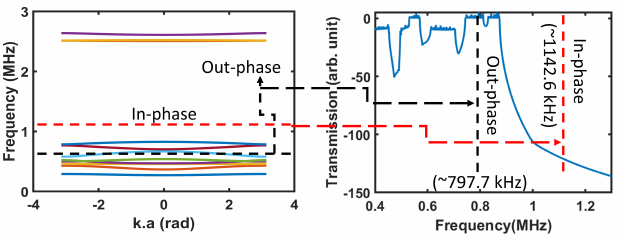}
    \caption{The simulated dispersion and transmission for the PnC that has a cavity resonator embedded, the out-phase mode of the cavity resonator lies in the band region, whereas the in-phase mode lies in the band gap.}
    \label{cavity_pnc}
\end{figure}

The embedded cavity resonator in the phononic crystal simulation reveals the locations of the in-phase and out-phase modes in both the dispersion curve and the transmission spectrum. As shown in fig. \ref{cavity_pnc}, the in-phase mode (marked in red) lies within the bandgap, whereas the out-phase mode (marked in black) lies outside the bandgap. This behavior demonstrates the ability of the phononic crystal support to localize wave energy within the cavity, reducing energy dissipation.


\begin{figure}[!h]
    \centering
    \includegraphics[width =\linewidth]{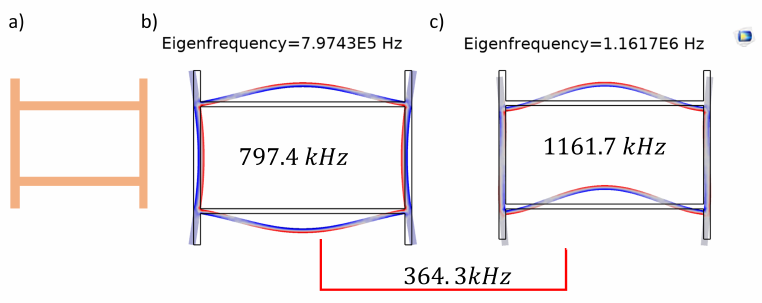}
    \caption{The frequency separation phenomena: a) the DETF cavity resonator without anchor or constraints, b) the simulated out-phase mode frequency, and c) the simulated in-phase frequency. The frequency difference of the simulated degenerate modes is 364.3 kHz, very similar to the experimental value.}
    \label{cavity_hypothesis}
\end{figure}

The cavity resonator supported by the PnC behaves like a structure without a conventional anchor. To demonstrate this, COMSOL simulations were performed on the cavity resonator without anchor, showing a frequency separation consistent with the experimental results. The experimentally measured frequency separation between the two phase modes is approximately 345 kHz, which agrees well with the simulation, as shown in fig. \ref{cavity_hypothesis}.

\bibliography{References}